\documentclass[journal]{IEEEtran}

\usepackage{cite}
\usepackage{amsmath,amssymb,amsfonts}
\usepackage{algorithmic}
\usepackage{graphicx}
\usepackage{textcomp}
\usepackage{algorithm}
\usepackage{algorithmic}
\usepackage{array}
\usepackage{color}
\usepackage{cite}
\usepackage{caption}
\usepackage{amsmath}
\usepackage{graphicx}
\usepackage[tight,footnotesize]{subfigure}
\usepackage{makecell}
\usepackage{float}
\usepackage{booktabs}
\usepackage{bm}
\usepackage{flushend}
\usepackage{tabularx}
\usepackage{enumerate}
\usepackage{multirow}
\usepackage{float}
\usepackage{amsmath}
\usepackage{lipsum}
\usepackage{amsthm} % 如果要使用proof语句就必须要引入这个包

\usepackage{amsmath,lipsum}
\usepackage{cuted}
\usepackage{stfloats}
\ifCLASSINFOpdf
\else
\fi
\begin{document}
\title{Air-Ground Collaborative Mobile Edge Computing: Architecture, Challenges, and Opportunities}
%Latency-Aware Task Selection and Scheduling in Air-Ground Reconnaissance System with Mobile Edge Computing}
%Latency-Aware Reconnaissance Task Selection and Scheduling in Energy-Constrained UAV-Assisted Mobile Edge Computing System

\author{
\IEEEauthorblockN{Zhen Qin, Hai Wang, Yuben Qu, Haipeng Dai, and Zhenhua Wei}
\thanks{Z. Qin and H. Wang are with the College of Communications Engineering, the Army Engineering University of PLA, China.}
\thanks{Y. Qu (corresponding author) is with the Department of Computer Science and Engineering, Shanghai Jiao Tong University, China.}
\thanks{H. Dai is with the Department of Computer Science and Technology, Nanjing University, China.}
\thanks{Z. Wei is with the Xi'an Research Institute of High Technology, China.}% <-this % stops a
%\thanks{C. Dong and Q. Wu are with the Key Laboratory of Dynamic Cognitive System of Electromagnetic Spectrum Space, Nanjing University of Aeronautics and Astronautics, China.}
%\thanks{Manuscript received xxxx, 2018; revised xxxx, 2018.}
}


\markboth{}%
{Shell \MakeLowercase{\textit{et al.}}: Bare Demo of IEEEtran.cls for IEEE Journals}



\maketitle

\begin{abstract}
By pushing computation, cache, and network control to the edge, mobile edge computing (MEC) is expected to play a leading role in fifth generation (5G) and future sixth generation (6G). Nevertheless, facing ubiquitous fast-growing computational demands, it is impossible for a single MEC paradigm to effectively support high-quality intelligent services at end user equipments (UEs). To address this issue, we propose an \underline{a}ir-\underline{g}round \underline{c}ollaborative \underline{MEC} (AGC-MEC) architecture in this article. The proposed AGC-MEC integrates all potentially available MEC servers within air and ground in the envisioned 6G, by a variety of collaborative ways to provide computation services at their best for UEs. Firstly, we introduce the AGC-MEC architecture and elaborate three typical use cases. Then, we discuss four main challenges in the AGC-MEC as well as their potential solutions. Next, we conduct a case study of collaborative service placement for AGC-MEC to validate the effectiveness of the proposed collaborative service placement strategy. Finally, we highlight several potential research directions of the AGC-MEC.
\end{abstract}

%\begin{IEEEkeywords}
%\end{IEEEkeywords}

% \fi
\IEEEpeerreviewmaketitle

\section{Introduction}\label{Introduction}
\IEEEPARstart{F}{ifth-generation} (5G) network has been deployed worldwide and commercially available in 2020, which offers many more functions than previous generations \cite{9144301}. However, with the advancement of smart devices and Internet of Things (IoT) technology, as well as diversified applications (\emph{e.g.,} smart city, mobile augmented reality, face recognition, and autonomous driving), 5G networks cannot completely meet future rapidly growing traffic demands. Accordingly, sixth-generation (6G) have attracted increasing attention from both industry and academia, which will be transformative and revolutionize the wireless evolution form. 6G network is expected to effectively support high-quality services and unlimited connectivity for a large number of intelligent devices \cite{9237460}. Meanwhile, it brings great challenges to the computing power of centralized data center and intelligent terminals. The traditional cloud computing cannot meet the requirements of massive data processing, and computing power will be transferred from the network core to the network edge. Mobile edge computing (MEC) \cite{8016573} is an emerging computing paradigm that can push mobile computing, cache, and network control to the edge in the close proximity of mobile user equipments (UEs). MEC is envisioned to play a leading role in 6G by operating as an intermediate layer that provides fast and localized data processing for many critical and resource-constrained applications \cite{mahmood2019six}.

With the help of MEC, computation-intensive and latency-sensitive tasks can be offloaded for remote execution, which can enhance the computing ability, and reduce energy consumption and latency. MEC servers are usually deployed in a fixed fashion at the ground base stations (BSs), wireless access points (APs), and roadside units (RSU). Nevertheless, such formed traditional terrestrial infrastructure-based MEC system has its limitations, which may not work in many critical applications, such as military, emergency relief, and disaster response. In addition, since terrestrial MEC servers lacks mobility, it cannot meet the computation and connectivity demands with the spatio-temporal dynamics. In contrast to the terrestrial MEC network, due to the Line-of-Sight (LoS) links, flexible deployment, and maneuverability, the aerial MEC network consisting of unmanned aerial vehicles (UAVs), airships, and balloons equipped with MEC servers might compensate those weaknesses. The aerial MEC network is also faced with many challenges such as the limited battery life. As a result, considering intelligent endogenous and ubiquitous computational power requirements of 6G networks, it is impossible for a single MEC paradigm to accomplish such a difficult task.
\newcommand{\tabincell}[2]{\begin{tabular}{@{}#1@{}}#2\end{tabular}}
\begin{table*}[t]\footnotesize
\renewcommand\arraystretch{1.5}
\setlength{\abovecaptionskip}{2pt}
\setlength{\belowcaptionskip}{0pt}
\centering
\caption{A comparison of different MEC paradigms in AGC-MEC.}
\begin{tabular}{|c|c|c|c|c|c|}
  \hline
  \textbf{MEC Paradigms} & \textbf{UAV}& \textbf{Airship/Balloon}&\textbf{BS/AP/RSU}&\textbf{Vehicle}&\textbf{Powerful Mobile User} \\
  \hline
  Location & Air & Air & Ground & Ground & Ground\\
  \hline
  Cost & Small &High &High & Medium & Small\\
  \hline
  Availability & Medium & High & High & Medium & Low\\
  \hline
  Reliability & Low & High & High & Medium & Low\\
  \hline
  Mobility & \tabincell{c}{Mobile \\(3D)} & \tabincell{c}{Quasi-Stationary \\(3D)}& \tabincell{c}{Static \\(2D)}& \tabincell{c}{Mobile \\(2D and Restricted)} & \tabincell{c}{Mobile \\(2D and Very Restricted)}\\
  \hline
  Energy Supply & Poor & Medium & Abundant & Medium & Poor\\
  \hline
  Coverage & Medium & Large & Large & Medium & Small\\
  \hline
  Computation Power& Weak$\sim$Medium & Medium$\sim$Strong & Strong & Medium & Very Weak\\
  \hline
  Communication Ability & Weak$\sim$Medium & Medium$\sim$Strong & Strong & Medium & Very Weak\\
  \hline
  Storage Capacity & Weak$\sim$Medium & Medium$\sim$Strong & Strong & Medium & Very Weak\\
  \hline
  %Scalability
  %Intelligence
\end{tabular}
\label{TAB12}
\end{table*}

\begin{figure*}[t]
\centering
  \subfigure[Collaborations between air and ground]{
  \label{fig_system_a}
  \minipage{0.66\textwidth}
  \includegraphics[width=\textwidth]{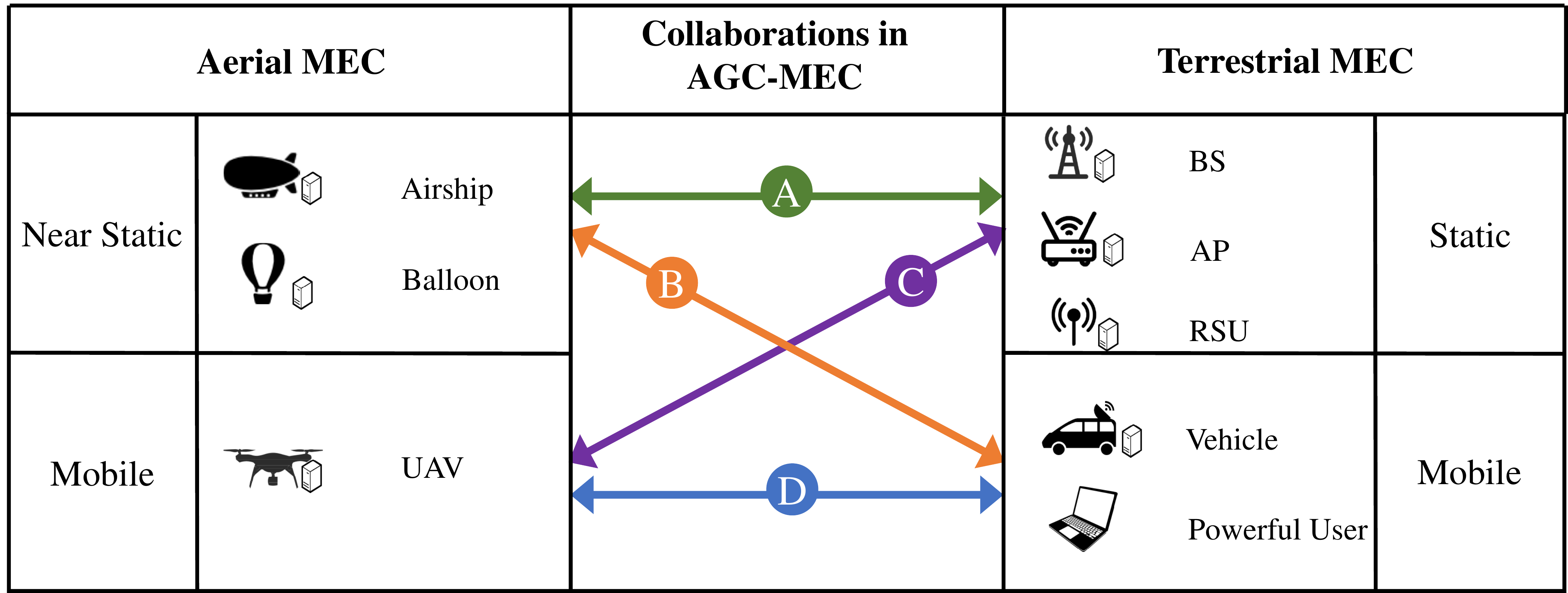}\\
  \endminipage
  }

  \subfigure[Typical use cases]{
  \label{fig_system_b}
  \minipage{0.7\textwidth}
  \hspace{1.2pt}
  \includegraphics[width=\linewidth]{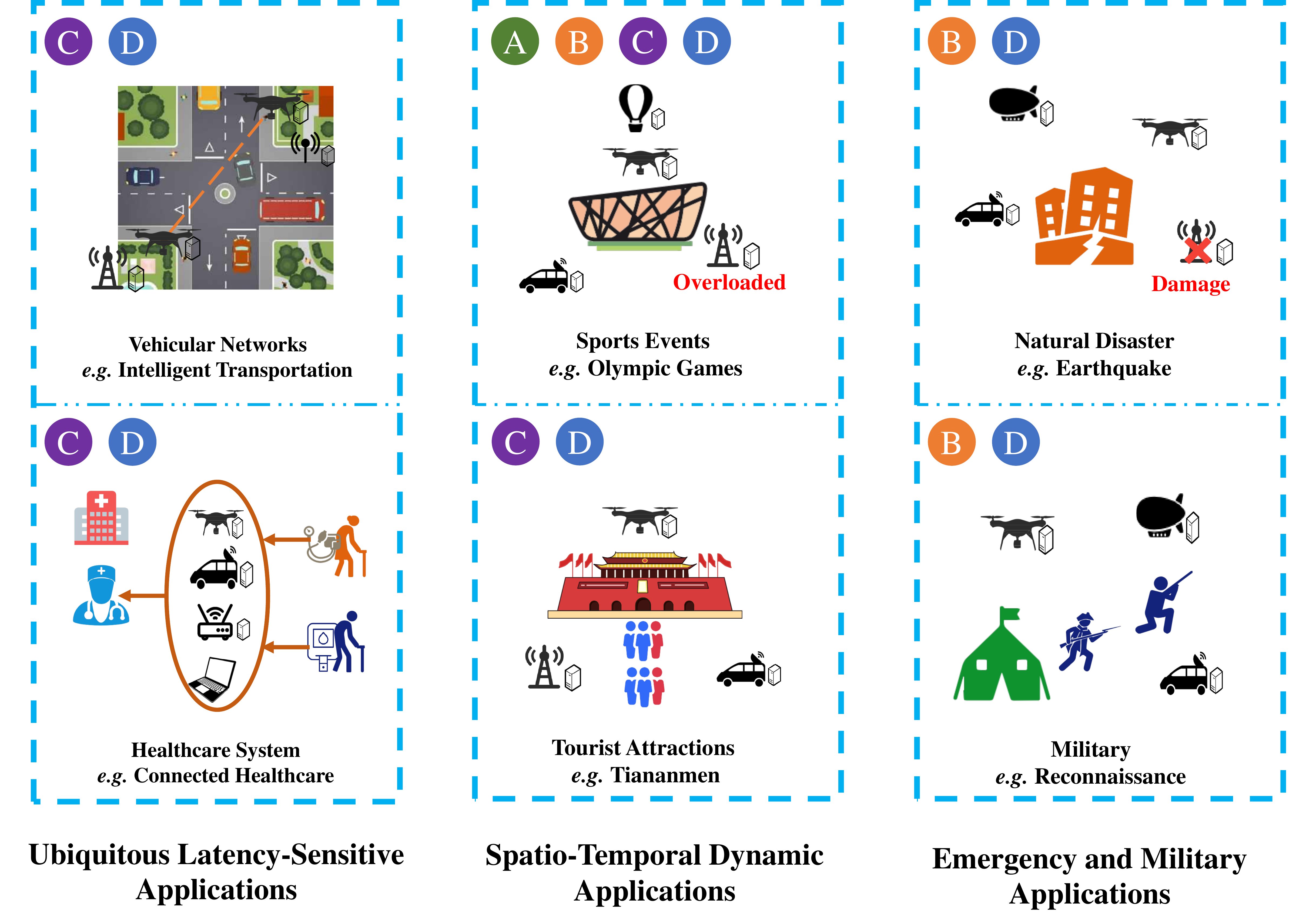}\\
  \endminipage
  }
\caption{AGC-MEC architecture.}
\label{fig_system}
\end{figure*}
To promote the development of 6G networks, we propose an \underline{a}ir-\underline{g}round \underline{c}ollaborative \underline{MEC} (AGC-MEC) architecture, which actively explores the complementary integration of computational powers from both air and ground segments to provide intelligent deployment and management of different MEC paradigms. The novelty of the proposed AGC-MEC lies in that it involves all potentially available MEC servers in the air-ground integrated networks. Specifically, the AGC-MEC architecture comprises a two-layer networking architecture: aerial MEC and terrestrial MEC. In the aerial MEC, airships and balloons are deployed as (near) static aerial MEC servers, while UAVs as mobile aerial MEC servers. In the terrestrial MEC, BSs/APs/RSUs are deployed as static terrestrial MEC servers, while vehicles as mobile terrestrial MEC servers. Meanwhile, powerful mobile users can be used as opportunistic mobile ground MEC servers. Through different cooperations between the air and ground, the AGC-MEC architecture is more flexible than the single terrestrial MEC network and more powerful than the single aerial MEC network.

There exist several studies that investigate air-ground integrated MEC network. They can be roughly divided into two categories: one considers aerial MEC servers only and the other considers both aerial and terrestrial MEC servers. For the former one, Zhou \emph{et al.} introduced three UAV-enabled MEC architectures, which improve computation performance and reduce execution latency by integrating UAV into MEC networks \cite{8961914}. Cheng \emph{et al.} proposed a novel air-ground integrated mobile edge network (AGMEN), where UAVs are flexibly deployed and scheduled, and assist the computing, caching, and communication of the edge network \cite{8436041}. For the latter one, Zhou \emph{et al.} proposed an air-ground integrated MEC framework, where ground vehicles and UAVs are envisaged as supplementary MEC servers for efficient service provisioning \cite{8436043}. Jiang \emph{et al.} proposed a heterogeneous MEC (H-MEC) architecture, which aims to address the key challenges of the H-MEC architecture in dynamic environments using AI-based solutions \cite{jiang_ai_2020}. However, all the aforementioned works only consider partial MEC paradigms in the air-ground integrated network. For example, \cite{8961914,8436041} consider the aerial MEC solely, while \cite{8436043} considers mobile ground vehicles and UAVs, and \cite{jiang_ai_2020} additionally considers the fixed ground BS based on \cite{8436043}. In contrast, our proposed AGC-MEC involves all potentially available MEC servers (both static and mobile) within air and ground in the context of future 6G networks. Furthermore, AGC-MEC considers how to enable different MEC servers collaborate effectively, rather than merely integrate them.

In the rest of this article, we first present the proposed AGC-MEC architecture and elaborate three typical use cases to illustrate its potential values in Section \ref{AGCMEC}. Then, we analyze several main technical challenges of the AGC-MEC in Section \ref{Challenges}. Next, we discuss three potential research directions in Section \ref{Directions}, while we conduct a case study to evaluate the performance of the AGC-MEC in Section \ref{Case}. Finally, we draw conclusions in Section \ref{Conclusion}.

\section{Air-Ground Collaborative Mobile Edge Computing (AGC-MEC)}\label{AGCMEC}
In this section, we propose the AGC-MEC architecture, which integrates all potentially available MEC servers within air and ground by a variety of collaborative ways to deliver computation services at their best for user equipments (UEs). Firstly, we introduce the AGC-MEC architecture, which consists of aerial and terrestrial MEC networks. Secondly, in order to show the potential of the AGC-MEC, we elaborate three typical use cases.

\subsection{AGC-MEC Overview}
Fig.~1~(a) illustrates the conceptual architecture of the proposed AGC-MEC, which comprises a two-layer networking architecture following the general air-ground integrated networks: aerial MEC and terrestrial MEC. Specifically, the aerial MEC network is mainly composed of UAVs, airships, and balloons, while the terrestrial MEC network is generally made up of BSs, APs, RSUs, vehicles, and mobile users, all of which are equipped with edge computing servers. In the aerial MEC, airships and balloons are deployed as (near) static aerial MEC servers and UAVs as mobile aerial MEC servers. In the terrestrial MEC, there are three types of MEC servers. Specifically, BSs/APs/RSUs and vehicles are employed as static and mobile terrestrial MEC servers, while some powerful mobile users can be used as opportunistic mobile terrestrial MEC servers. We envision that all the aforementioned MEC servers can collaborate with each other in the proposed AGC-MEC, which mainly focuses on the collaborations between air and ground. As shown in Fig.~1~(a), the collaborations can be divided into four categories as follows: A) static-air and static-ground collaboration; B) static-air and mobile-ground collaboration; C) mobile-air and static-ground collaboration; D) mobile-air and mobile-ground collaboration. In practical, several collaborations could work together.

Different MEC paradigms have their particular features, which are partially overlapping but also complementary. We present a comprehensive comparison of different MEC paradigms in the AGC-MEC in Tab.~1. MEC servers are usually deployed in a fixed fashion at the BSs, APs, RSUs, airships, and balloons. Despite their high availability and reliability, they also bring higher costs than UAVs, vehicles and powerful mobile user. Due to mobility and flexibility, UAVs, vehicles can be deployed quickly on demand. UAVs move much faster than vehicles, but with less computation, communication and storage resources. On the contrary, vehicles move slower but hold more resources. In fact, BSs, APs, RSUs, airships and balloons have the most available resources. Furthermore, the coverage of fixed MEC server is large but remains unchanged. They cannot exploit its mobility to move closer to UEs with computation-intensive tasks. In contrast, UAVs, vehicles and powerful mobile users have limited coverage and energy, but can move close to UEs to provide low-latency services and communication. By collaborating different MEC paradigms of air and ground, AGC-MEC aims to manage and control heterogeneous network resources smartly to meet computing demands.

\subsection{Typical Use Cases}
As illustrated in Fig.~1~(b), the proposed AGC-MEC architecture is envisioned to be useful particularly in several applications as follows.
\subsubsection{Ubiquitous Latency-Sensitive Applications}
With the development of IoTs and various mobile applications, more and more data is generated at the edge of the network. It is a general trend to process and analyze the data in real time widely at the network edge, which needs strong computing power support. By effectively collaborating different MEC paradigms, the AGC-MEC architecture can provide efficient and flexible computing services at the edge to meet the demands of ubiquitous latency-sensitive applications. For example, intelligent transportation system (ITS) requires low-latency communication and high computation capabilities. Static ground RSUs equipped with MEC servers can process the local data, which not only reduces the burden of network transmission, but also speeds up the data processing speed. Nevertheless, there are situations where additional air and ground MEC servers are required to handle temporary high traffic loads during extreme traffic congestion or unexpected weather conditions. According to congestion conditions and traffic events, UAVs can be dynamically deployed as mobile aerial MEC servers. In addition, MEC can play a significant role in connected healthcare systems by offering better insight of heterogeneous healthcare content to support affordable and quality patient care \cite{li_edgelaas_2019}. Ubiquitous collaborative MEC servers can help patients choose better advice from the right guardians in real time when some emergencies occur.

\subsubsection{Spatio-Temporal Dynamic Applications}
In the daily life, mobile UEs have obvious group effect and usually form different dense crowds with bursty requests over time, which results in a spatio-temporal dynamic computing demand. The AGC-MEC can deploy various MEC servers on demand to meet such a dynamic demand of UEs. One typical example scenario is the stadium in a major sport or concert event. There are a huge number of people gathered, who execute computation-intensive applications in their mobile phones such as Virtual Reality (VR) and online gaming. In this case, BSs may be overloaded and cannot support massive UEs. Fortunately, UAVs, airships, balloons and vehicles can be temporary deployed to collaborate with ground BSs to offload computation tasks and improve the user quality of experience (QoE). Other typical example scenario is tourist attractions and important transportation hubs. For example, on October 3, 2019, Tiananmen received 2.68 million tourists throughout the day; the passenger flow reached the peak of that day at the time of flag lowering, which was 180 thousand. To meet the dynamic computing demands of tourists, UAVs can be deployed flexibly to collaborate with ground MEC systems during holidays.

\subsubsection{Emergency and Military Applications}
In some extreme cases such as emergency and military applications, the ground infrastructure-based MEC system may be destroyed partially or completely and thus cannot work properly. The AGC-MEC can assemble mobile MEC servers including UAVs, airships, balloons, and vehicles to assist the existing communication infrastructure, if any, more critically, in providing computing services. For instance, in the event of natural disasters and earthquakes, ground infrastructure may be damaged. The rescue crews may need mobile augmented reality to search the area, which needs a significant amount of computing resources. The AGC-MEC can provide the required computation resources to increase the search scope and speed up the rescue. Moreover, in military applications, there are massive computation-intensive reconnaissance tasks (\emph{e.g.,} estimating the locations and the dynamics of the hostile forces). For a large-scale reconnaissance area without infrastructure, UAVs can collaborate with ground vehicles to expand the reconnaissance scope, reduce cloud computing latency, and enhance the computing ability.

\section{Challenges in AGC-MEC}\label{Challenges}
Due to the features of high mobility, heterogeneity, frequent inevitable air-ground interactions and time-varying channel conditions, the AGC-MEC architecture is difficult in platform integration, network deployment, resource management, and intelligence realization. In this section, we discuss four challenges and their potential solutions.
\subsection{Generic Computing Platform Integration}
Generic computing platform integration is critical to ensure synergy between different MEC paradigms. The AGC-MEC involves different MEC paradigms, which exist in different segments and have distinct characteristics. MEC paradigms are connected to each other through a communication protocol and communicate with users and devices. However, different MEC paradigms have various communication protocols, communication links, and interfaces, which significantly limits interoperability \cite{8436041}. Therefore, how to integrate the computing resources of various MEC paradigms and build a generic computing platform is our primary consideration.

Network Function Virtualization (NFV) is an emerging network technology, which moves the network function from the original special equipment to the general equipment. Specifically, it decouples network functions from specialized hardware, and can be leveraged to flexibly implement network functions as software instances in the network slices. Despite its great potential benefits, NFV is also faced with some problems to be solved. For example, under the virtualized network environment, which interfaces can be used privately, and which interfaces need to be standardized, these issues remain to be clarified.

\subsection{On-Demand 3D Network Deployment}
In the AGC-MEC architecture, a fundamental and critical issue is how to deploy air/ground MEC servers, which includes two aspects. For one thing, based on the demand, we need to determine when, where, what MEC paradigms and how many MEC servers to deploy. For another thing, we need to determine how to collaborate optimize the trajectories of mobile air/ground MEC servers.

There are many challenges for the on-demand deployment of edge servers. First, due to the high mobility of users, heterogeneity of QoS requirements, and stochastic characteristics of wireless network, the deployment of MEC servers is exposed to extreme difficulties. Second, since AGC-MEC architecture involves the cooperation between air and ground, the MEC servers need to be deployed in 3D space. The critical design challenge is to adaptively adjust the trajectories to meet the dynamic computing task requirements. The existing work considers to use Deep Reinforcement Learning (DRL) algorithm, which can learn optimal placement policies and plan trajectories of mobile MEC servers intelligently \cite{jiang_ai_2020}. However, DRL algorithm brings more computational cost than traditional methods.

%\subsection{Computing Powers Collaboration Management}
%The AGCMEC architecture should not only provide powerful computing powers to handle complex tasks, but also provide an efficient collaboration management. Effective collaboration can enable various MEC paradigms to cooperate with each other with high QoS guarantee. Due to heterogeneity and mobility, the collaboration among air and ground is very challenging. A key issue to be solved is who controls and coordinates various edge computing servers. One way is centralized control, but there is a single point of failure (SPOF) problem. Meanwhile, it will bring immense overheads to control center so that additional computing cannot be afforded. The other way is distributed control, which means each MEC servers collaborative with their neighbors distributively. Although distributed control solves SPOF problem, it brings higher inaccuracy \cite{zhou_air-ground_2018}. Hybrid control has the characteristics of the above two control modes at the same time. Part of the control functions run on the centralized controller, and part of the functions are distributed on other edge computing servers.

\subsection{Computing-Oriented Resource Allocation}
Resource allocation is an important guarantee for effective collaboration of AGC-MEC architecture, which directly impacts network performance. In order to effectively complete computing tasks, the AGC-MEC architecture should carry out efficient, adaptive, and intelligent resource allocation. Air-ground collaborative resource allocation involves various resources, which can be mainly divided into the following two types: the communication resource including bandwidth and channels, and the computation resource including computing power and storage capacity. In particular, to provide personal services for different users, MEC servers need to store corresponding data including object databases, libraries and trained machine learning models, associated with services. Service provisioning is an essential and critical issue, \emph{i.e.,} how to determine where to store/place which MEC service to meet various computing service demands. Furthermore, jointly optimizing resource allocation, trajectory, and placement of MEC servers can obtain globally optimal performance that would be more useful in practice.

There are two main challenges to deal with the above problem. First, the problem includes many decision variables to be jointly optimized, \emph{e.g.,} the continuous computation resource allocation and UAV trajectory variables, and integral task offloading and service placement variables. Therefore, the optimization problem is a non-convex mixed integer nonlinear programming (MINLP) problem and is hard to solve in general. Secondly, computing demands vary in time and space. In order to match resource provision with computing tasks, resource allocation needs to be adjusted dynamically in a real-time manner. In order to better adapt to the dynamic environment, the complicated network behaviors can be analyzed in real time, and the online algorithm can be used to flexibly adjust resource allocation.
\subsection{Ubiquitous Edge Intelligence Realization}
Edge intelligence (EI) is emerging as a promising key enabler for MEC to fulfill the vision of ubiquitous intelligence, which pushes intelligence to the network edge by running artificial intelligence (AI) algorithms on edge devices. EI can be divided into two main types of technology: AI for edge and AI on edge \cite{9052677}. The former focuses on utilizing AI algorithms to provide effective solutions for key problems in edge computing, and the latter focuses on realizing AI model training and inference on the edge. However, it is difficult to realize ubiquitous EI in the following aspects.

\textbf{AI for AGC-MEC:} It is devoted to provide better solutions to constrained optimization problems, \emph{e.g.,} trajectory optimization, resource allocation, network deployment, and real-time decision making. Despite its great potential benefits, utilizing AI algorithms is also faced with some problems to be solved. Firstly, since some MEC servers is resource-constrained, how to balance optimality and efficiency of the AGC-MEC architecture is a great challenge. Secondly, if we want to utilize AI algorithms to obtain solutions, the formulated optimization problem and mathematical model need to be restricted \cite{9052677}. Therefore, the model establishment is a huge challenge.

\textbf{AI on AGC-MEC:} The computing and storage capacity of edge servers is far less than that of cloud servers, which cannot meet the needs of a large number of computing and storage resources for AI training. Fortunately, federated learning (FL) can make multiple resource-constrained end devices to collaboratively train effective learning models, which is an emerging distributed learning architecture. However, it is challenging in learning-oriented training configuration, and energy efficient training strategies.

\section{Case Study: Collaborative Service Placement for AGC-MEC}\label{Case}
In this section, we conduct a case study of collaborative service placement for AGC-MEC to validate the effectiveness of the proposed collaborative service placement strategy. Firstly, we introduce the system model and problem formulation. Secondly, we describe the simulation settings and comparison algorithms. Finally, we show the simulation results.

\subsection{System Model and Problem Formulation}
\subsubsection{Network Model}
We consider an air-ground collaborative MEC network, which is composed of one UAV, one BS, and multiple UEs. The UAV and BS can collaborative with each other to provide various types of computing services for UEs. The computation-intensive tasks of UEs can be operated locally or be offloaded to the UAV or BS.

\subsubsection{Latency Model}
The latency is composed of two parts: computation time taken to execute tasks, communication time taken  to offload tasks. The time for transmitting computation results is usually ignored. The computation time is calculated by the required number of CPU frequency cycles and allocated computing resources. The communication time is calculated by the input data size and transmission rate.
\subsubsection{Energy Model}
The energy consumption of the UE includes computation and communication energy consumption. The computation energy is related to the effective switched capacitance, local computing power and required number of CPU frequency cycles. The communication energy is related to the communication time and transmission power of the UE.
\begin{figure*}[!htb]
%8910
\label{task}
\centering
\subfigure[Convergence result]
{
    \label{fig:a}
    \includegraphics[width=0.75\columnwidth]{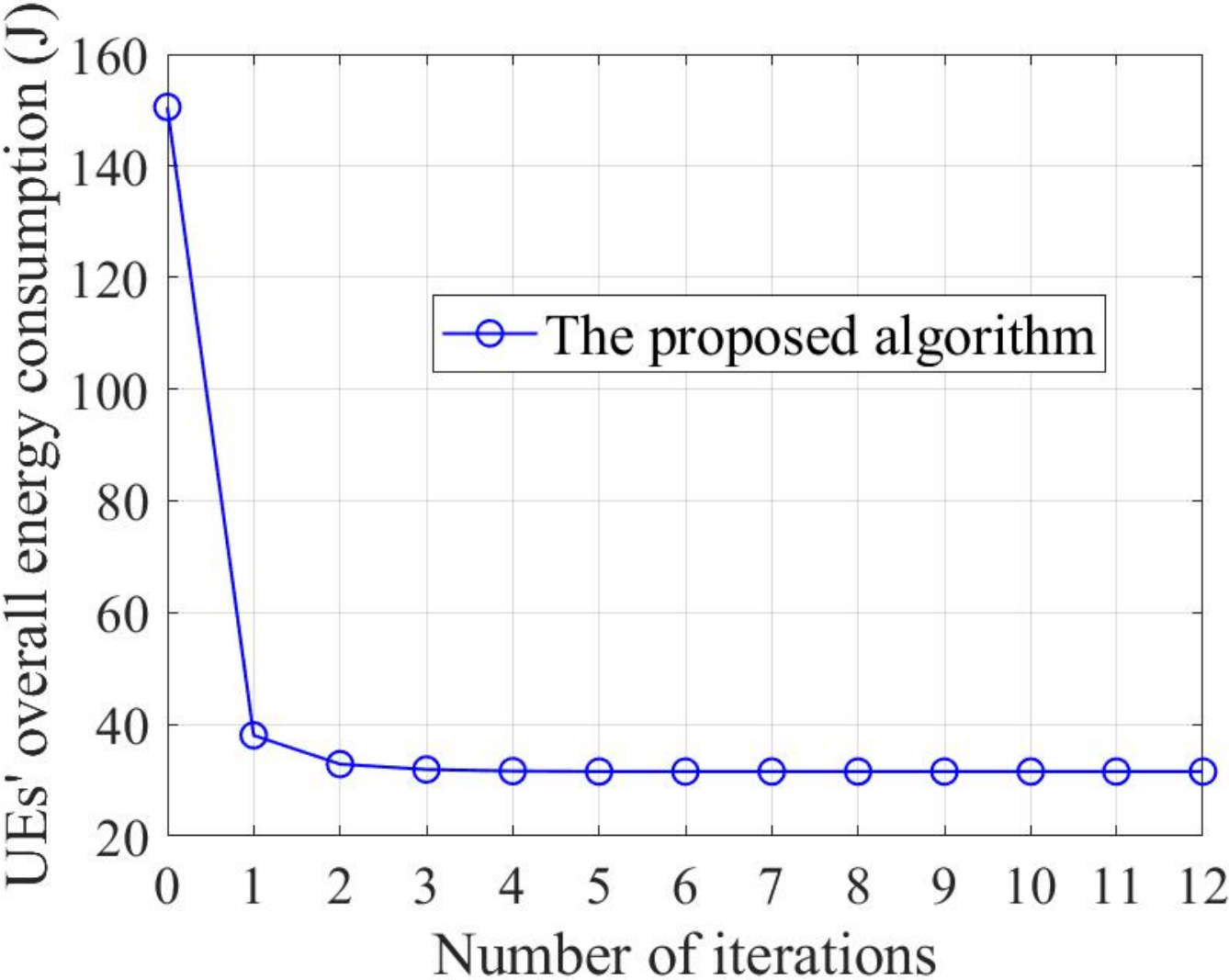}
}
%\hspace{-0.3in}
\subfigure[UAV trajectory]
{
   \label{fig:b}
    \includegraphics[width=0.75\columnwidth]{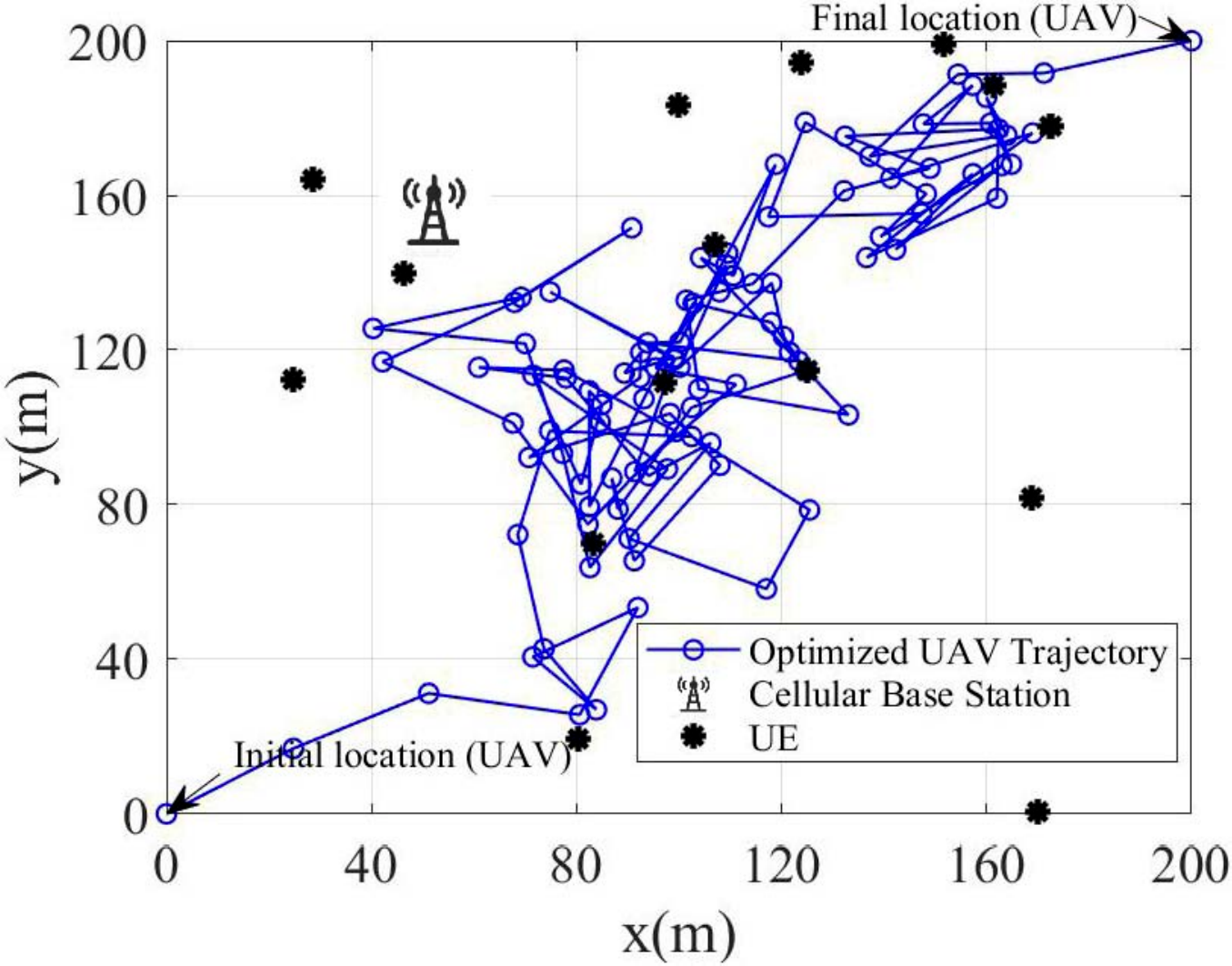}
}
%\hspace{-0.3in}
\caption{Convergence and optimized UAV trajectory.}
\end{figure*}
\subsubsection{Problem Formulation}
We aim to minimize UEs' overall energy consumption by jointly optimizing service placement, task offloading, UAV trajectory and computation resource allocation. The optimization problem can be formulated as:
\begin{itemize}
\item Optimization objective: the minimization of UEs' overall energy consumption.
\item Optimization variables: service placement, task offloading, UAV trajectory and computation resource allocation.
\item Constraint 1: storage capacities of the BS and UAV, which store corresponding data including object databases, trained machine learning models, and libraries, associated with services.
\item Constraint 2: the restriction of maximum number of UEs associated with the BS and UAV.
\item Constraint 3: the coverage area of the UAV.
\item Constraint 4: the limitation of flight distance.
\item Constraint 5: offloading condition, \emph{i.e.,} the required service need to be placed in the BS or UAV.
\end{itemize}

\begin{figure*}[!htb]
%8910
\label{task}
\centering
\subfigure[Overall energy consumption \emph{v.s.}  UAV's storage capacity ]
{
    \label{fig:a}
    \includegraphics[width=0.75\columnwidth]{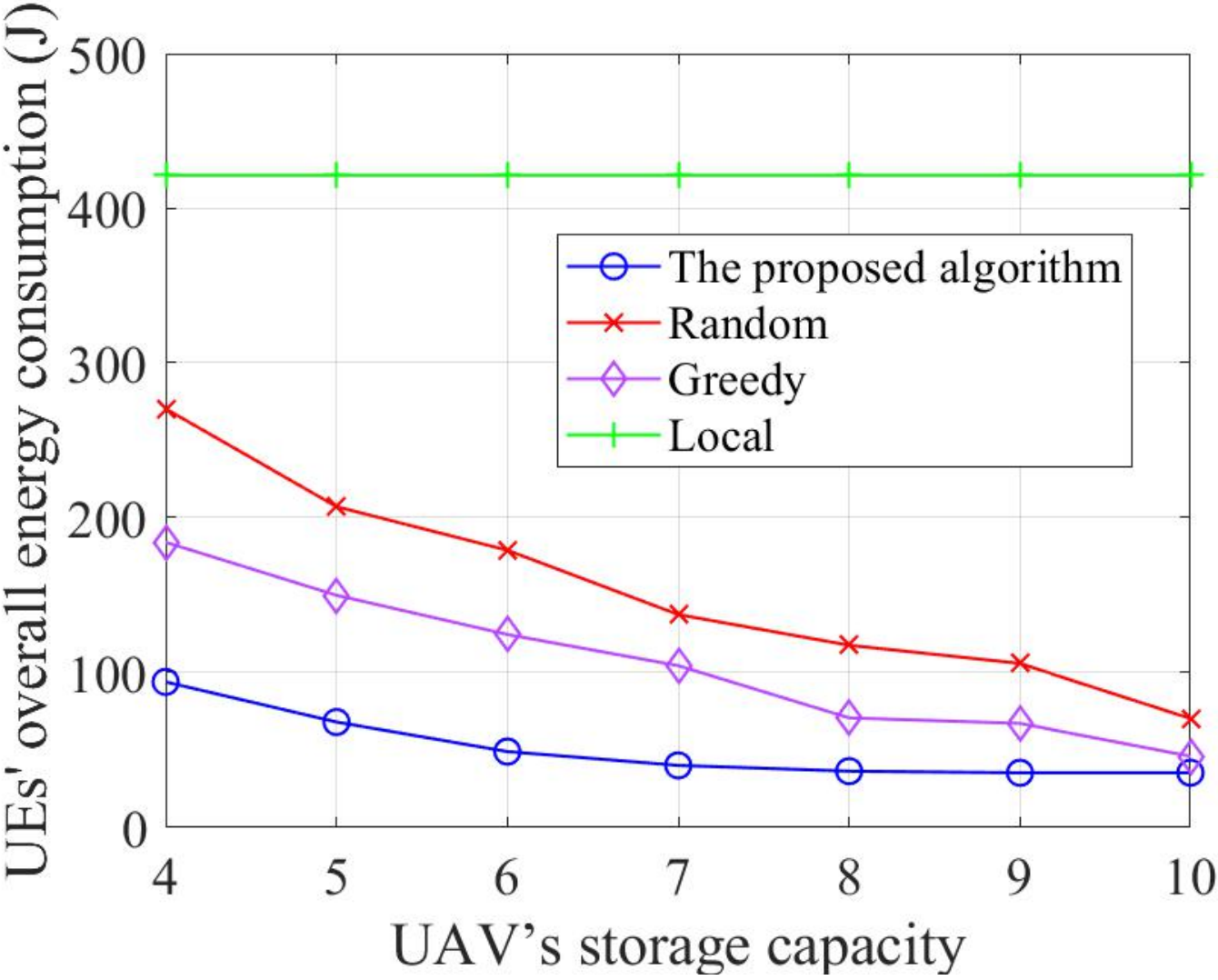}
}
%\hspace{-0.3in}
\subfigure[Overall energy consumption \emph{v.s.} UEs' workload]
{
   \label{fig:b}
    \includegraphics[width=0.75\columnwidth]{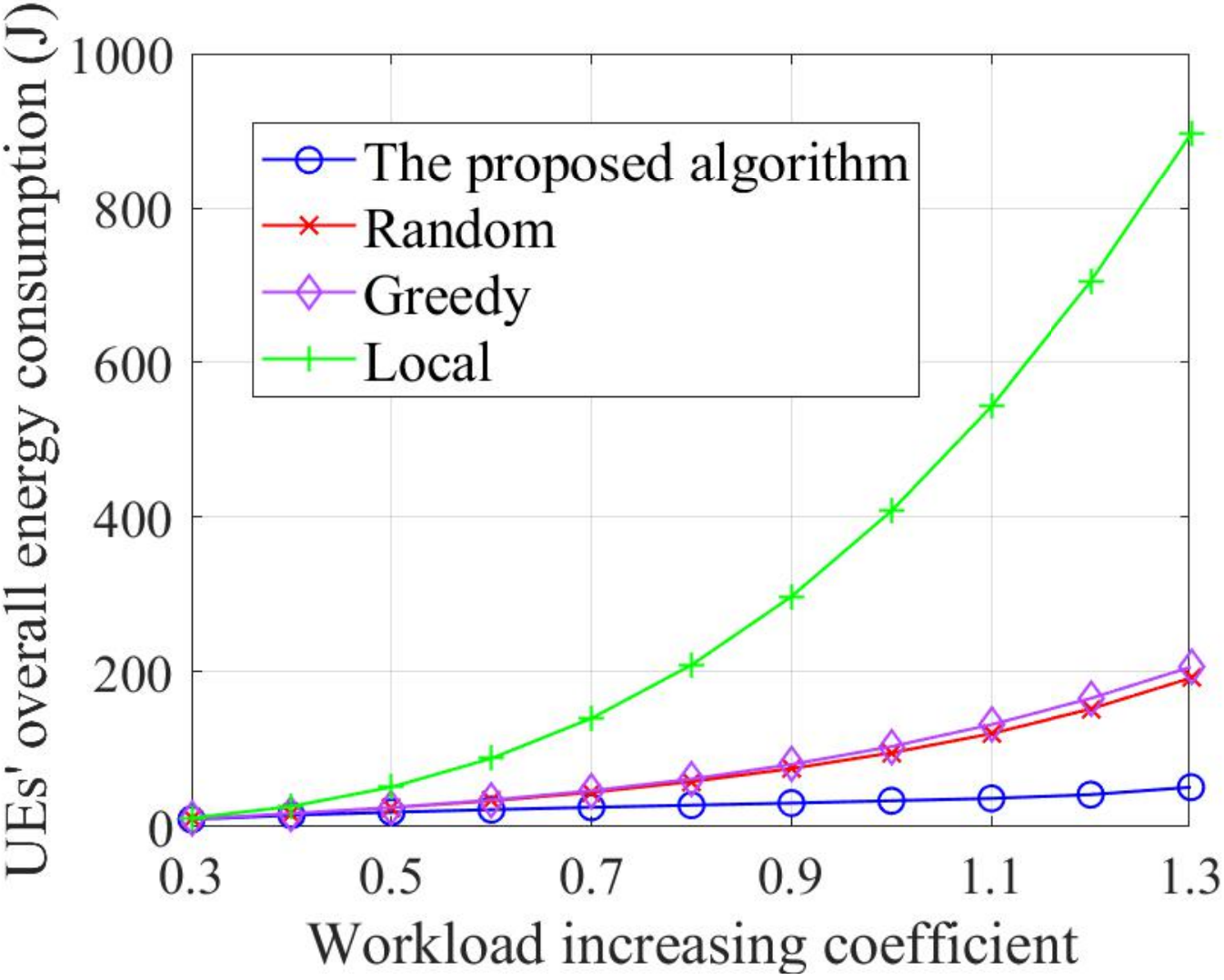}
}
%\hspace{-0.3in}
\caption{Effects of UAV's storage capacity and UEs' workload.}
\end{figure*}

The optimization problem is a MINLP problem, which includes continuous computation resource allocation and UAV trajectory variables and integral task offloading and service placement variables. To deal with this problem, we exploit alternating optimization techniques to propose a suboptimal solution with convergence guarantee. To be specific, we obtain the closed form of the optimal computation resource allocation and iteratively solve the task offloading and service placement subproblem by Branch and Bound (BnB), and UAV trajectory subproblem by successive convex approximation (SCA).

\subsection{Simulation Settings and Comparison Algorithms}
This case considers an air-ground collaborative MEC network with a mobile UAV, a static BS and ten UEs in a ${\rm{200m}} \times {\rm{200m}}$ squared area. The UAV and BS are equipped with MEC servers and UEs are randomly distributed in the squared zone. We divide the whole mission period into 100 time slots, and the length of time slot is $1s$. Each UE generates a computation task constantly in each time slot. The number of required CPU cycles is in the range of ${\rm{[1}}{{\rm{0}}^{\rm{8}}},{10^9}]$ cycles and the size of input data is in the range of ${\rm{[100}},1000]$KB. We assume that the service population of UEs' requests are produced according to the popular Zipf distribution with a skewness parameter value of $0.5$ form a service set with size $30$ \cite{8737368}. The storage size required by each services is uniformly chosen from $[0.5,1]$. We assume that the storage capacity of the BS is twice that of the UAV. The maximum transmission power of UEs equals to $0.1w $ and the bandwidth is $1MHz$. For computation energy consumption model, we assume the effective switched capacitance is ${10^{27}}$. UAV related parameters are set as follows: the maximum ground coverage radius is $100m$, maximum flying distance in a time slot is $30m$, maximum computation capacity is ${\rm{[5}},10]GHz$, and the maximum number of UEs associated with UAV is ${\rm{[3}},5]$.

To verify the performance of the proposed algorithm, we compare it with the following three algorithms:
\begin{itemize}
\item Random: according to storage capacities of the BS or UAV, this algorithm places services randomly;
\item Greedy: this algorithm uses a greedy strategy to select the service with the minimum required storage, which results in maximizing the number of services placed at the BS or UAV;
\item Local: this algorithm executes tasks locally, which obtains an upper bound of UEs' overall energy consumption.
\end{itemize}

\subsection{Simulation Results}
Fig. 2 (a) shows the convergence performance of the proposed algorithm. It can be seen that after three iterations, UEs' overall energy consumption decreases from 150.5956 J to 31.8719 J, which means the proposed algorithm converges quickly. Fig. 2 (b) demonstrates the trajectory of the UAV optimized by the proposed algorithm, where the initial and final horizontal positions of the UAV are (0,0) and (200,200), respectively. As shown in Fig. 2 (b), the proposed algorithm enables the UAV to fly close to UEs. This is because that the UAV can provide better computing service in the close proximity of UEs.

Fig. 3 (a) shows the trend of UEs' overall energy consumption under different UAV's storage capacity. Meanwhile, the storage capacity of the BS is also changing, which is twice that of the UAV. With the increase of storage capacity, UEs' overall energy consumption by all algorithms except Local has been decreased. This is because, the UAV and BS can store more MEC services required by UEs with larger storage capacity. Compared with Random, Greedy and Local, UEs' overall energy consumption of the proposed algorithm can be reduced by 65.96\%, 49.35\%, 87.89\%, respectively. Fig. 3 (b) demonstrates the trend of UEs' overall energy consumption under different UEs' workload (\emph{i.e.,} the number of required CPU cycles). In this simulation, we use the workload increasing coefficient to represent the change of workload. As shown in Fig. 3 (b), compared with Random, Greedy and Local, UEs' overall energy consumption of the proposed algorithm can be reduced by 46.62\%, 48.42\%, 75.53\%, respectively.

\section{Potential Research Directions of AGC-MEC}\label{Directions}
In despite of its great potential, the study on the AGC-MEC architecture is still in its infancy, where many critical problems should be solved. In this section, we introduce three potential research directions that can help translate the visions of the AGC-MEC into reality.
%\subsection{Common/Open Architectures}
\subsection{Network Control}
As previously mentioned, the AGC-MEC network involves different MEC paradigms belonging to different segments distributed in a wide area and thus needs to be well managed. As is known to all, software defined networking (SDN) separating the data plane and control plane introduces a unified control plane interface and global view of the whole network. SDN is able to provide centralized network control and flexible resource management of the collaboration among different MEC paradigms. Still, the SDN-based AGC-MEC is faced with many critical issues. For example, the placement of SDN controllers is a key to the SDN-based AGC-MEC, which should take many factors into consideration such as energy efficiency, load balancing, latency, scalability, \emph{etc} \cite{8802245}.

\subsection{Security and Privacy}
Since MEC servers in the AGC-MEC are located at the network edge and have frequent interactions, they lack effective backup and recovery measures of their data, which is prone to be attacked or misused by malicious users. Furthermore, compared with the cloud computing data center in the core network, MEC can collect more high-value sensitive information of UEs, including location information, lifestyle, social relations, even health status, \emph{etc.} Therefore, the security and privacy protection is critical to the AGC-MEC architecture. Fortunately, Blockchain is a burgeoning distributed ledger technology, which has the potential to ensure the data and resource exchange among untrusted MEC nodes being safe in the AGC-MEC.

\subsection{Intelligent Collaboration}
For a long-lasting sophisticated mission, it may require multiple MEC servers to provide services together. To be specific, some MEC servers may act as controllers to determine offloading decisions. According to the decisions, some MEC servers may be used as relays to offload task to servers with strong computing power and rich resources. And then these MEC servers may cooperate to provide computing services. Therefore, it is valuable to study how many MEC servers are needed to complete task and how to intelligently collaborate.

\section{Conclusion}\label{Conclusion}
In this article, we have proposed the AGC-MEC architecture to explore the complementary integration of all potentially available MEC servers within air and ground by various collaborative ways to provide high-quality intelligent services for future 6G network. We have described the AGC-MEC architecture and three typical use cases. The challenging issues and their potential solutions have also been discussed. Furthermore, we have conducted a case study of collaborative service placement for AGC-MEC, and presented several potential research directions for future study.

%\section*{Acknowledgment}
%\begin{thebibliography}{00}

%\bibitem{b1}Handbook of Unmanned Aerial Vehicles, K. P. Valavanis and G. J. Vachtsevanos, Springer Netherlands, 2015.
%\bibitem{b2}I. Bekmezci, O. K. Sahingoz, and S. Temel, ``Flying ad-hoc networks (FANETs): A survey,'' \textit{Ad Hoc Networks}, vol. 11, no. 3, pp. 1254-1270, May. 2013.
%\bibitem{b3}R. Singh, M. Thompson, S. A. Mathews, O. Agbogidi, K. Bhadane, and K. Namuduri, ``Aerial base stations for enabling cellular communications during emergency situation,'' in \textit{Proceedings of IEEE International Conference on Vision, Image and Signal Processing}, 2017, pp. 103-108.
%\bibitem{b4}B. Hament, and P. Oh, ``Unmanned aerial and ground vehicle (UAV-UGV) system prototype for civil infrastructure missions,'' in \textit{Proceedings of IEEE International Conference on Consumer Electronics}, 2018, pp. 1-4.
%\bibitem{b5}C. Barrado, R. Messeguer, J. Lopez, E. Pastor, E. Santamaria, and P. Royo, ``Wildfire monitoring using a mixed air-ground mobile network,'' \textit{IEEE Pervasive Computing}, vol. 9, no. 4, pp. 24-32, Oct. 2010.

%\end{thebibliography}
%This work is supported in part by National NSF of China
%under Grant No.61472445, No.61702525 and No.61702545, in part by the NSF of Jiangsu
%Province under Grant No. BK20140076.
\appendices
\bibliographystyle{IEEEtran}
\bibliography{main}

\ifCLASSOPTIONcaptionsoff
  \newpage
\fi

%\newpage

\clearpage

\end{document}